\documentclass[aps,prd,epsf]{revtex4}

\begin{document} 

\title{
A new look at anomaly cancellation in heterotic $M$-theory} 
\author{Ian G. Moss}
\email{ian.moss@ncl.ac.uk}
\affiliation{School of Mathematics and Statistics, University of  
Newcastle Upon Tyne, NE1 7RU, UK}

\date{\today}


\begin{abstract}
This paper considers anomaly cancellation for eleven-dimensional supergravity
on a manifold with boundary and theories related to heterotic $M$-theory. The
Green-Schwarz mechanism is implemented without
introducing distributions. The importance of the supersymmetry anomaly in
constructing the low energy action is discussed and it is argued that a
recently proposed action for heterotic $M$-theory gives a supersymmetric theory
to all orders in the gravitational coupling $\kappa$.
\end{abstract}

\maketitle
\section{Introduction}

Horava and Witten have argued that the strong coupling limit of the
ten-dimensional heterotic string is eleven-dimensional supergravity with gauge
multiplets confined to two ten-dimensional hypersurfaces forming the boundary
of the eleven-dimensional spacetime manifold 
\cite{horava96,horava96-2}. 
This theory is a very promising starting point for phenomenological models
based on compactifications to four dimensions (see, for example,
\cite{witten96,banks96,lukas98,lukas98-2,ellis99}). In applications such as
these, it is important to know the action in as much detail as possible.

The form of the action originally put forward was found by relying partly on
anomaly cancellation and supersymmetry. Gauge and gravitational anomalies in
the theory
cancel via a novel modification of the Green-Schwarz mechanism involving the
supergravity 3-form. The cancellation, which involves some remarkable
algebraic coincidences, requires that the matter action contains a factor of
order $\kappa^{2/3}$ compared to the supergravity action, where $\kappa$ is
the eleven-dimensional gravitational coupling strength.

Imposing local supersymmetry on the action fixes all of the terms at order
$\kappa^{2/3}$. However, when the same procedure is applied to order
$\kappa^{4/3}$, singular terms depending on the square of the delta-function
start to arise. This problem has recently been overcome by modifying the
boundary conditions on the gravitino and the supergravity 3-form, so that now
an action can be constructed which is non-singular and supersymmetric to
higher orders \cite{moss03,Moss:2004ck}. The effect of these boundary
conditions on anomaly cancellation is one of the issues to be addressed in
this paper.

As we extend the theory to higher orders in the gravitational coupling, we have
to take account of the supersymmetry anomaly which appears at order $\kappa^2$
(i.e. $\kappa^2$ times the gravitational action). The existence of a
supersymmetry anomaly implies that the classical action should not be
supersymmetric at this order. However, it is reasonable to suppose that the
supersymmetry anomaly, like the gauge anomaly, is cancelled by the
Green-Schwarz mechanism, and the action should therefore be supersymmetric up
to the variation of the Green-Schwarz terms
\cite{Itoyama:1985ni,Itoyama:1985qi}. This was not appreciated in
\cite{Moss:2004ck}, where it was shown that the supersymmetric variation of
the new action for heterotic $M$-theory reduced to a single term of precisely
this type. Now, taking into account the supersymmetry anomaly, this theory
appears to be supersymmetric to all orders in $\kappa$, at least when truncated
to terms up to first order in the Riemann tensor.

A heuristic argument for cancellation of the supersymmetry anomaly by the
Green-Schwarz terms can be made from the Wess-Zumino consistency conditions,
which  relate the supersymmetry anomaly to the gauge anomaly 
\cite{Itoyama:1985ni,Itoyama:1985qi}. 
When the Green-Schwarz terms are added to the effective action, the total has
vanishing gauge variation. Therefore, provided that the consistency conditions
have a unique solution, the variation of the Green-Schwarz terms should cancel
the supersymmetry anomaly as well.

We shall consider the gauge and supersymmetry anomaly cancellation in more
detail. We work throughout on the `downstairs picture' of a manifold with
boundary, rather than lifting to the covering space $R^{10}\times S_1$.
For the present, we truncate the action to first order in the
Riemann tensor.  The gauge anomaly from the chiral fermion on one of the
boundary components can be described by a formal 12-form $ I_{12}(F)$
\cite{green}. To
generate the anomaly, we introduce the notation $T$, such that locally
$dT\omega=\omega$ for a closed form $\omega$. The anomalous variation of the
chiral fermion effective action under gauge transformations
$\delta_\alpha$ is given by integrating a 10-form $I^A_{10}$, defined by
\begin{equation}
I^A_{10}=T\delta_{\alpha}TI_{12}
\end{equation}
The anomalous variation under supersymmetry variations is given by the sum of
two other 10-forms, $ I^S_{10}+ I^{S'}_{10}$, where according to 
\cite{Itoyama:1985ni,Itoyama:1985qi},
\begin{equation}
I^S_{10}=l_\eta TI_{12}
\end{equation}
and $I^{S'}_{10}$ is gauge invariant. The anti-derivative $l_\eta$ is defined
by $l_\eta A=0$ and $l_\eta F=\delta_\eta A$.

In the case of the gauge group $E_8$, $(4\pi)^5I_{12}=({\rm tr}F^2)^3/12$ and
we have
\begin{eqnarray}
I^A_{10}&=&{1\over 12(4\pi)^5}{\rm tr}(\alpha F)( {\rm tr}F^2)^2\label{ia}\\
I^S_{10}&=&{1\over 12(4\pi)^5}{\rm tr}(\delta_\eta A A)( {\rm tr}F^2)^2
+{1\over 3(4\pi)^5}{\rm tr}(\delta_\eta A F)T( {\rm tr}F^2)^2.\label{is}
\end{eqnarray}
Now the observation of Horava and Witten was that $I^A_{10}$ can be
cancelled by a variation of the $CGG$ term in the supergravity action
\cite{cremmer78}. This can be done by requiring $G\sim {\rm tr}F^2$ on the
boundary and $\delta_\alpha C\sim\delta(x^{11}){\rm tr}(\alpha F)$.
If we follow this route further, we are eventually lead to the theory with
$\delta(x^{11})^2$ terms \cite{horava96-2}.

An alternative way to arrange the Green-Schwarz cancellation was first
described in \cite{deAlwis:1996hr}. If we
let $\delta_\alpha C\sim da$, where $a$ is any
2-form which satisfies $a={\rm tr}(\alpha F)$ on the boundary, and require that
$G\sim {\rm tr}F^2$ on the boundary, then the
variation of the Green-Schwarz term is a total derivative,
\begin{equation}
\delta CGG\sim d(aGG).
\end{equation}
This integrates to give a term which can cancel the anomaly  (\ref{ia}).
Similarly, if we add an extra supersymmetry variation $\delta_\eta C\sim df$
to the 3-form, where $f={\rm tr}(\delta_\eta A A)$ on the boundary, then part
of the supersymmetry anomaly $I^S_{10}$ is cancelled \cite{Moss:2004ck}. (The
rest of the anomaly is cancelled by the usual transformation of $C$, as we
shall see shortly).

The gauge and supersymmetry variations of $C$ are precisely those which are
required to maintain the gauge and supersymmetry invariance of the 3-form
boundary condition given in \cite{Moss:2004ck}, namely
\begin{equation}
C=\frac{\sqrt{2}}{12}\epsilon\,\left(\omega_{3Y}
+\omega_\chi\right)\label{bcc}
\end{equation}
where $\omega_{3Y}$ is the Chern-Simons form $T{\rm tr}F^2$ and
\begin{equation}
\omega_\chi=\frac14\bar\chi^a\Gamma_{ABC}\chi^a.\label{chichi}
\end{equation}
The constant $\epsilon$ is fixed by supersymmetry to the normalisation of the
matter action. It is related to the gauge coupling $\lambda$ by
$\epsilon=\kappa^2/2\lambda^2$. 
 Since the Chern-Simons form has a gauge transformation
$\delta_\alpha\omega_{3Y}=d({\rm tr}\,\alpha F)$, the boundary condition
remains valid if the variation of $C$ is  given by
\begin{equation}
\delta_\alpha C={\sqrt{2}\over 12}\epsilon\, da\label{ctrans}
\end{equation}
where $a={\rm tr}(\alpha F)$ on the boundary. (Some details of the use of
$p$-form boundary conditions in quantum field theory can be found in 
\cite{Moss:1990yq,Moss:1994jj}. A more careful treatment would consider
the Abelian BRST variations of the boundary condition, but these are similar in
form to the Abelian gauge variations.)

The fermion term (\ref{chichi}) in the boundary condition is required to make
the boundary condition supersymmetric. It also plays an important role in
obtaining the correct ten-dimensional reduction. Unfortunately, the gaugino
enters into the variation of the $CGG$ term through the value of $G=6dC$ on
the boundary,
\begin{equation}
G=\frac{\epsilon}{\sqrt{2}}\left( {\rm tr}\,F^2+d\omega_\chi\right).
\end{equation}
In order to avoid spoiling the anomaly cancellation, we have to
add extra boundary corrections to the Green-Schwarz terms. The $CGG$ term is
taken from the usual supergravity action (with gravitational coupling
$\kappa^2/2$ \cite{Conrad:1997ww}),
\begin{equation}
S_C=-{2\sqrt{2}\over \kappa^2}\int_{\cal M} CGG.
\end{equation} 
The boundary terms can only involve $\omega_{3Y}$, $\omega_\chi$ and $F$ and
they must vanish when $\omega_\chi=0$. The unique combination which has the
desired effect is
\begin{equation}
S_3=-{\epsilon^3\over6\kappa^2}\int_{\partial \cal M}\omega_{3Y}\
\omega_\chi
\,\left(2\,{\rm tr}F^2+d\omega_\chi\right).
\end{equation}
The variation of $S_C$ and $S_3$ using (\ref{ctrans}) is
\begin{equation}
\delta_\alpha S_C+\delta_\alpha S_3=
-{\epsilon^3\over6\kappa^2}\int_{\partial \cal M}
{\rm tr}(\alpha F)( {\rm tr}F^2)^2.
\end{equation}
The variation cancels the gauge anomaly (\ref{ia}) and fixes the value of
$\epsilon$,
\begin{equation}
\epsilon={1\over 4\pi}\left({\kappa\over 4\sqrt{2}\pi}\right)^{2/3}.
\end{equation}
This agrees with \cite{Conrad:1997ww}, which corrected a factor of $2$ in
\cite{horava96-2}. The result differs by a factor of $3$ from the one obtained
on the covering space in \cite{Bilal:1999ig}. The difference is possibly due
to the way in which the theory is lifted to the covering space. 

If our assumptions are correct, then the supersymmetric variation of the 
Green-Schwarz terms should now cancel the supersymmetry anomaly,
\begin{equation}
\delta_\eta S_C+\delta_\eta S_3+\int ( I^S_{10}+I^{S'}_{10})=0.
\end{equation}
A supersymmetry variation of $S_C$ and $S_3$ allows us to read off the
non-gauge-invariant part
of the anomaly
\begin{equation}
I^S_{10}={\epsilon^3\over 6\kappa^2}
{\rm tr}(\delta_\eta A\,A)({\rm tr}F^2)^2+
{2\epsilon^3\over 3\kappa^2}{\rm tr}(\delta_\eta A F)\omega_{3Y}{\rm
tr}F^2.
\label{susyngi}
\end{equation}
This is in complete agreement with (\ref{is}), proving that this part of the
supersymmetry anomaly does indeed cancel. We also generate the gauge invariant
part of the supersymmetry anomaly,
\begin{equation}
I^{S'}_{10}={\epsilon^3\over \kappa^2}
{\rm tr}(\delta_\eta A F)\omega_\chi\left(2\,{\rm tr}F^2+d\omega_\chi\right)
+{\epsilon^3\over 6\kappa^2}
(\delta_\eta\omega_\chi)\omega_\chi \left(3\,{\rm tr}F^2+2d\omega_\chi\right).
\label{susygi}
\end{equation}
In these expressions, we have local supersymmetry transformations
\begin{eqnarray}
\delta_\eta A&=&\frac32\bar\eta\Gamma_A\chi\\
\delta_\eta \omega_\chi&=&
\frac18\bar\eta\Gamma_{ABC}\Gamma^{DE}\chi^a\hat F^a{}_{DE}
+\frac38\bar\eta\Gamma_D\psi_{[A}\bar\chi^a\Gamma^D{}_{BC]}\chi^a,
\end{eqnarray}
where $\hat F_{AB}=F_{AB}-\bar\psi_{[A}\Gamma_{B]}\chi$.
The supersymmetry anomaly in ten dimensions has been calculated previously up
to four fermi terms \cite{Grignani:1987mi,Fisher:1987kq}. Our
result has a similar form, although a direct comparison is not worthwhile
because our result is specific to the gauge group $E_8$ and contains
contributions from the eleventh dimension (indicated by the presence of the
gravitino field $\psi_A$).

We can make use of the anomaly (\ref{susyngi})  in connection with the action
of heterotic $M$-theory. The action $S$ proposed in \cite{Moss:2004ck}
consisted of usual supergravity action  with boundary terms $S_0$ and a
boundary matter action
\begin{equation}
S_1=-{2\epsilon\over\kappa^2}\int_{\cal \partial M}dv
\left(\frac14{F^a}_{AB}
{F^a}^{AB}+\frac12\bar\chi^a\Gamma^AD_A(\Omega^{**})\chi^a
+\frac14\bar\psi_A\Gamma^{BC}\Gamma^A{F^a}^*_{BC}\chi^a\right),
\label{action1}
\end{equation}
where $F^*=(F+\hat F)/2$, $\Omega$ is the supergravity spin connection and
$
\Omega^{**}{}_{ABC}=\Omega_{ABC}+\frac1{24}\psi^D\Gamma_{ABCDE}\psi^{E}.
$
We have discovered a new result that we must also add the term
$S_3$ for the anomaly cancellation to work properly. In \cite{Moss:2004ck}, it
was shown that the supersymmetric variation of the action was
\begin{equation}
\delta_\eta S={2\sqrt{2}\over\kappa^2}\int_{\cal \partial M}
\delta_\eta C \,C \,G,
\end{equation}
up to one possible four fermi term and all orders in $\kappa$. We now
recognise
this as the variation of the Green-Schwarz term, and therefore cancells with
the supersymmetry anomaly. The extra four fermi terms in (\ref{susygi})
explain also why there was a four-fermi term left in the variation. Up to the
limitations of truncating out the higher order curvature terms, the action
$S=S_0+S_1+S_3$ describes a theory which is supersymmetric to all orders in
$\kappa$.

The treatment of gauge and gravitational anomalies  in the original
Horava-Witten model included terms  which are higher order in the Riemann
tensor \cite{horava96-2}. The 12-form which generates the anomalies was
obtained from the
gauginos and boundary effects on the gravitino \cite{horava96},
\begin{equation}
I_{12}={1\over 12(2\pi)^5}\left(I_4^3-4I_4X_8\right),
\end{equation}
where $I_4={\rm tr}F^2-\frac12{\rm tr}R^2$ and
$X_8=-\frac18{\rm tr}R^4+\frac1{32}({\rm tr}R^2)^2$.
The gauge, gravitational and supersymmetry anomalies due to the first term in
$I_{12}$ can be removed by the $CGG$ term as described above, provided that the
boundary condition on $C$ is modified,
\begin{equation}
C=\frac{\sqrt{2}}{12}\epsilon\,\left(\omega_{3Y}-\frac12\omega_{3L}
+\omega_\chi-\frac12\omega_\psi\right),\label{bccr}
\end{equation}
where $\omega_{3L}=T{\rm tr}R^2$ and $\omega_\psi$, according to dimensional
analysis,  is bilinear in the derivative of the gravitino. 
The construction of a fully supersymmetric theory with this boundary condition
has not yet been done, but it seems inevitable that $R^2$ terms will also
appear in the action \cite{Lukas:1997fg,Lukas:1998ew}. These terms would be
needed to ascertain the precise form of $\omega_\psi$.

Similarly, the $X_8$ terms in the gauge anomaly should be cancelled by a
Green-Schwarz term $CX_8$ in eleven dimensions 
\cite{horava96-2,Lukas:1998ew}. 
This can be done with $\delta_\alpha C\sim da$ as above, but the cancellation
is not exact, because there are extra terms in the eleven dimensional
curvature appearing in the Green-Scwartz term which are not present in the ten
dimensional curvature part of the anomally. These would
be removed by adding additional boundary terms, or possibly by finding new
contributions to the anomaly.

In conclusion, it is possible to cancel the gauge anomalies in eleven
dimensional supergravity with boundaries without introducing
singular gauge transformations. The $CGG$ term in the supergravity action acts
as a Green-Schwarz term, but with fermions present in the boundary conditions 
it is necessary to introduce an extra boundary term depending on the gaugino
field. It is interesting that the boundary conditions and action appear to be
well-determined from gauge and supersymmetry invariance without making any use
of the covering space. This agrees with recent work by van Nieuwenhuizen and
Vassilesvich \cite{vanNieuwenhuizen:2005kg}, who have found that supersymmetry
severely restricts  the boundary conditions for pure supergravity. Given also
that eleven dimensional supergravity with more than two boundaries can now be
consistently formulated (at least as $\kappa\to0$) \cite{Freed:2004yc}, it
looks increasingly likely that the manifold with boundary picture is the more
fundamental way of formulating heterotic $M$-theory.

We have seen the supersymmetry anomaly has to be taken into account when
constructing the action and, at least in the limit of small curvature, there
is an action for supergravity with boundary matter which gives a fully
supersymmetric quantum field theory. However, dimensional reduction to four
dimensions involves curvature terms in the internal dimensions which are not
small, but comparable in size to the gauge field strength
\cite{witten96,banks96,lukas98,lukas98-2,ellis99}. 
It would be very desirable to find a supersymmetric action which includes the
$R^2$ terms suggested by the gauge and gravity anomalies, and then we would
have confidence in using the theory as a basis for particle physics
phenomenology.

\bibliography{super.bib,books.bib}


\end{document}